%% file: GGS1_preprint.tex
\documentclass[5p,twocolumns]{elsarticle}
\usepackage[latin1]{inputenc}
\usepackage[english]{babel}
\usepackage{graphicx}
\usepackage{epsfig}
\usepackage{subfigure}
\usepackage{hhline}
\usepackage{amssymb,amsmath}
\usepackage[]{natbib}
\usepackage{eurosym} 
\addto\captionsenglish{  }
\begin{document}
\newcommand{\Angst}{$\mathring{\mathrm{A}}$}
\newcommand{\DEG}{$^\circ$}
\newcommand{\sq}{$^\mathrm{2}$}

\begin{frontmatter}

\title{Generic guide concepts for the European Spallation Source}
\author[ess]{C. Zendler\corref{cor1}}
\ead{carolin.zendler@esss.se, ph: +46-721792232}
\author[ess]{D. Martin Rodriguez}
\author[ess,uppsala]{P. M. Bentley}
\address[ess]{European Spallation Source ESS AB, Box 176, 221 00 Lund, Sweden}
\address[uppsala]{University of Uppsala, Uppsala, 751 20, Sweden}
\cortext[cor1]{Correspondig author}

\begin{abstract}
  The construction of the European Spallation Source (ESS) faces many challenges from the neutron beam transport point of view: The spallation source is specified as being driven by a 5 MW beam of protons, each with 2 GeV energy, and yet the requirements in instrument background suppression relative to measured signal
  vary between 10$^{-6}$ and 10$^{-8}$.  The energetic particles, particularly above 20 MeV, which are expected to be produced in abundance in the target, have to be filtered in order to make the beamlines safe, operational and provide good quality measurements with low background.

  We present generic neutron guides of short and medium length instruments which are optimized for good performance at minimal cost. Direct line of sight to the source is avoided twice, with either the first point out of line of sight or both being inside the bunker (20\,m) to minimize shielding costs. These guide geometries are regarded as a baseline to define standards for instruments to be constructed at ESS. They are used to find commonalities and develop principles and solutions for common problems. Lastly, we report the impact of employing the over-illumination concept to mitigate losses from random misalignment passively, and that over-illumination should be used sparingly in key locations to be effective.  For more widespread alignment issues, a more direct, active approach is likely to be needed.
\end{abstract}


\end{frontmatter}

\section{Introduction}

Ground breaking at the ESS construction site took place in September 2014 just outside the city of Lund, Sweden. 
At the time of writing, three of the 22 planned public instruments have already entered the detailed design phase of the construction project, with more following every year.
With such a large number of instruments designed by different partners at almost the same time, it is important to identify commonalities and find cost-effective solutions to common problems in order to avoid duplication of effort and associated cost increases. Moreover, a baseline guide concept for each instrument category is a useful tool to benchmark new ideas and guide concepts. 

The design of guides for a source with the characteristics of ESS has unprecedented challenges. The first challenge is the adequate geometry for efficient neutron transport over distances as long as 150\,m. So far, beamlines like the high resolution powder diffractometer HRPD~\cite{HRPD} at ISIS with a length of 100\,m count amongst the longest instruments. This challenge has been addressed extensively in several studies (e.g. \cite{Kleno,Stahn2011S12,Bertelsen2013387,EG2}) by the use of ballistic or elliptic guides which minimize reflection losses.
The second challenge is to reduce the extent of beam losses due to misalignments. This problem has not been addressed as extensively as the first challenge, and needs to be evaluated.
The third challenge, and by no means the least important, is the appropriate geometry to obtain low background while maintaining high transport efficiency. Beamlines between 20\,m and 50-75\,m face an unprecedented challenge, due to the high proton beam intensity, which places design requirements to mitigate the spallation background risk that may contaminate the useful neutron beam. This background poses a problem not only for the measurements aspiring a high signal to noise ratio, but also affects the safety design of ESS, as well as mitigation of radiation damage for components, activation, and the amount of shielding required as a consequence along the beamline.

The requirements on neutron background are particularly stringent for SANS, reflectometry, and spectroscopy.  The requirements call for a noise suppression relative to the signal of around 10$^{-6}$ and 10$^{-8}$ in the strongest cases --- see, for example, \cite{HERITAGE_PROPOSAL}.  Such backgrounds are possible to achieve at spallation sources \cite{REFLECTOMETRY_TS2}, but only with careful optimisation of the whole system of optics and shielding together.  In this article, for brevity we concentrate on the optical part of the problem, and only describe some of the shielding in much broader terms.

Where baseline beamline shielding designs from neighbouring beamlines overlap, a common shielding area has been defined that the ESS calls the bunker.  This structure is similar to other guide bunkers at existing facilities.  Taking advantage of this bunker, by losing line of sight before the beamline emerges into the guide hall is one way that individual instrument costs can be reduced.  Equipment in direct view of the source is illuminated by stray hadrons\footnote{Mostly neutrons, but also some protons and pions} spanning the MeV to GeV energy range, and these produce showers of secondary particles (mostly neutrons), therefore a direct view of any such secondary source has to be avoided as well.  That is to say, line of sight to the source should be avoided twice.  For direct geometry spectrometers, the concept of double crystal monochromators \cite{HybridSpectrometer} naturally fulfills this condition. All other instrument classes need alternative solutions.

The magnitude of the aforementioned challenges depend on the length of the guide to some extent.  It is currently anticipated that, with longer neutron guides, it will be possible to reduce the background by taking advantage of distance, and the ESS is currently examining fast neutron albedo transport.   Presently, for the long guides the technical focus is on maintaining low guide costs and minimising the transport losses from misalignment.  In contrast, the problem is the opposite for the short beamlines, because the misalignment losses are less significant and the total guide costs are lower. However, the line of sight avoidance condition is severely more restrictive, and the design of a guide with a good performance and a low background becomes more of a challenge.

In this article, after establishing a basic technical grounding, we will examine medium length neutron guides of 50\,m, in section~\ref{sec_Generic50m}, including both the double line of sight requirement as well as a study of misalignment, before solutions for 20\,m long instruments focussing on the line of sight condition are discussed in section~\ref{sec_Generic20m}.

\section{Simulation details}

\label{sec_TechnicalStuff}

\subsection{Simulation program}

For all simulations, the Monte-Carlo ray-tracing package VITESS~\cite{Vitess,Vitess3} version 3.2 is used. Guide cross-sections are rectangular and gravitational effects are included by default.
The neutron source that was used in the models is the ESS TDR moderator\footnote{VITESS moderator characteristics of the \texttt{2013\_Schoenfeldt} database with a 12$\times$12\,cm\sq\,moderator size}~\cite{TDR}.
More recent moderator developments do not change the validity of the present work, since line of sight is avoided in the horizontal direction. An adoption to a 3\,cm high flat moderator (like the so-called ``butterfly'' which is currently planned for ESS) can easily be done by a modification of the vertical guide shape to reoptimise the vertical plane beam extraction. The difference in the horizontal geometry that the new moderators bring are most important for bi-spectral instruments, and --- since this only affects the first sections of guide traversing the monolith bulk shielding, amounting to a small perturbation overall --- these are not considered here.
The viewable area of the moderator surface in the horizontal plane is approximately about 7\,cm (cold) and 12\,cm (thermal) --- these are larger than the effective areas that couple to the guide, and large compared to the 2\,cm beam desired for a chopper around the 6\,m position (see section~\ref{sec_Generic50m} below), therefore only a slight adjustment of the beam extraction would be needed to adapt the 50\,m instrument solution to the new moderator design. Similarly, the short instrument optimization starts altogether at 6\,m from the source  (see section~\ref{sec_Generic20m} below), after a simple straight monolith insert that can be adjusted independently if need be, but is with a width of 3\,cm small enough to not see any difference between a 7\,cm and a 12\,cm wide moderator. 

Supermirror reflectivity is modelled by a generalized non-linear description~\cite{HenriksBispectral,Vitess3}. Non-linear guide shapes are modelled using straight mirror segments.

\paragraph{Optimization}
The main optimization routine used here is the particle swarm optimization (PSO), as available within the VITESS package since version 3.2, which follows the time-varying acceleration coefficients method \cite{Swarm_TVAC}.

\subsection{Line of sight and shielding standards}

In a curved neutron guide, if the radius necessary to avoid direct line of sight is calculated such that the line is exactly closed by the supermirror planes, particles along the critical trajectory traverse an infinitessimally small volume of guide substrate.  
Several instruments have been built with this design only to relearn the lesson that fast neutrons travel easily through glass and are not reflected by supermirrors, and those instruments suffer from background problems as a result.
Therefore, the radius $R$ of a curved guide is calculated assuming a width $w'$ larger than the separation of the supermirrors ($w$) by 5\,mm. In addition, the use of metal substrate guides feeding through large collimation blocks (sometimes called ``horse-collars'' at other facilities) in at least three parts of the curved section is planned (cf fig.~\ref{f_LoS_skizze}).

This is the baseline design until a minimum path length within shielding material is defined, by detailed shielding calculations.

\section{Double-ballistic neutron guides for medium length instruments (50\,m)}
\label{sec_Generic50m}

Several neutron guides are planned to be deployed at ESS as part of
``medium length'' instruments, corresponding to a guide length of
about 50\,m-75\,m. For reasons of background reduction, as well as
radiation safety --- which essentially translate directly into
instrument performance and cost --- the sample and detector position
should wherever possible \emph{not} have a direct line of sight to the
moderator.  This is the ``first line of sight'' principle.

Furthermore, to avoid illumination by secondary particles, the sample
and detector position should also avoid having a direct line of sight
view of any equipment that lies within line of sight of the source.
This is our ``twice out of line of sight'' recommendation.  It is
further advantageous for shielding cost reduction
\cite{NOSG_REVIEW_COST_OPTIMISATION_2014} to be out of line of sight
once within the common shielding bunker, which is foreseen to have a
radius of 20\,m for short instruments, and 30\,m for medium and long
instruments at the time of writing.

For the purposes of this study, the guide is designed with the following objectives:
\begin{itemize}
\item deliver high flux on a 1$\times$1\,cm\sq\,sample at 50\,m from
  source
\item avoid direct line-of-sight twice
\item homogeneous beam divergence
\item 2\,cm wide slit for fast choppers at 6.25\,m
\item 4\,cm wide slit for choppers somewhere around half distance
\end{itemize}
These objectives capture the typical, essential features of ESS beamlines,
and we study this as a generic concept not directly related to any
particular instrument.

\begin{figure}[!h]
\subfigure[]{\includegraphics[width=0.6\linewidth]{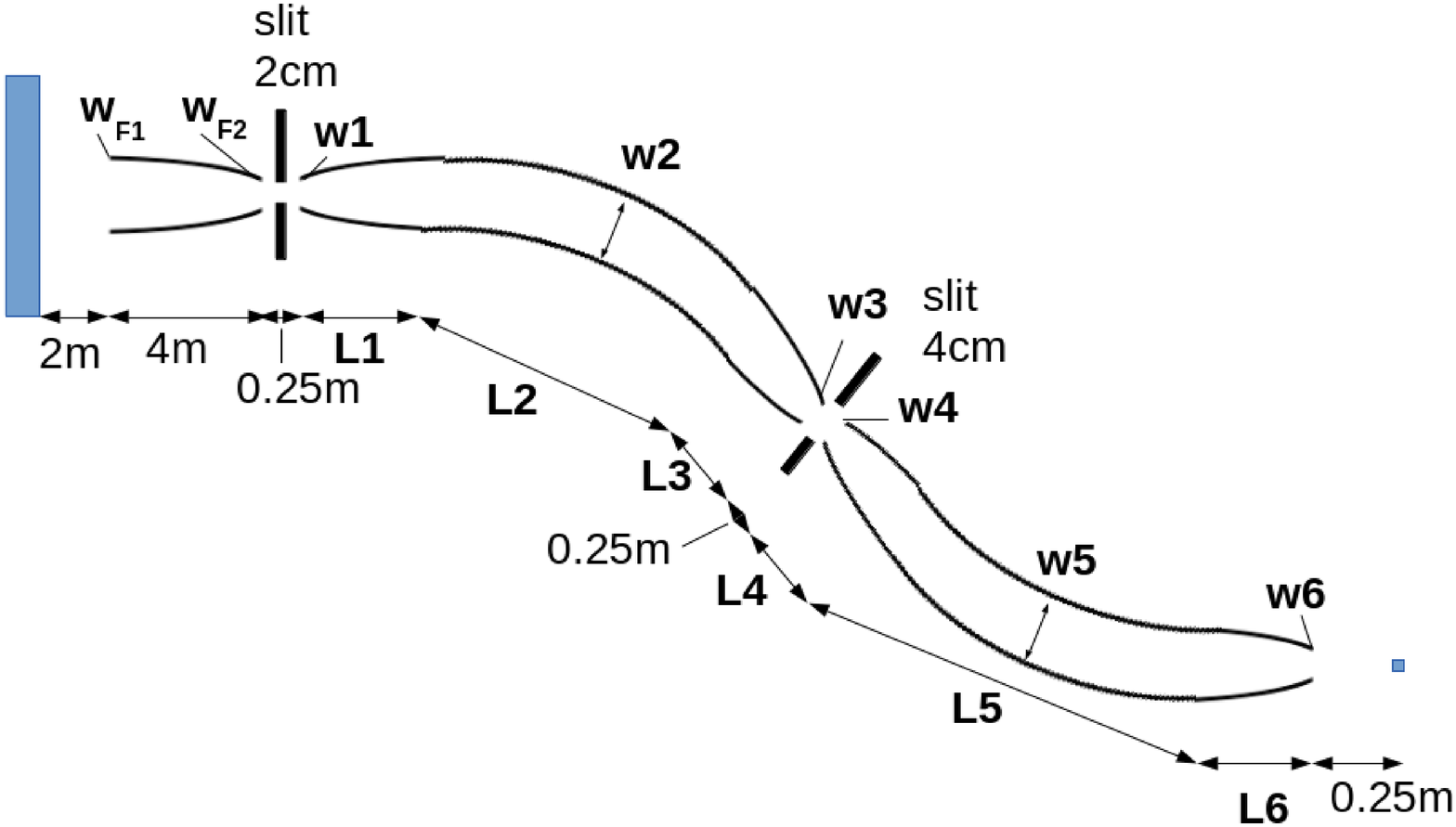}}
\subfigure[ \label{f_LoS_skizze}]{\includegraphics[width=0.25\linewidth]{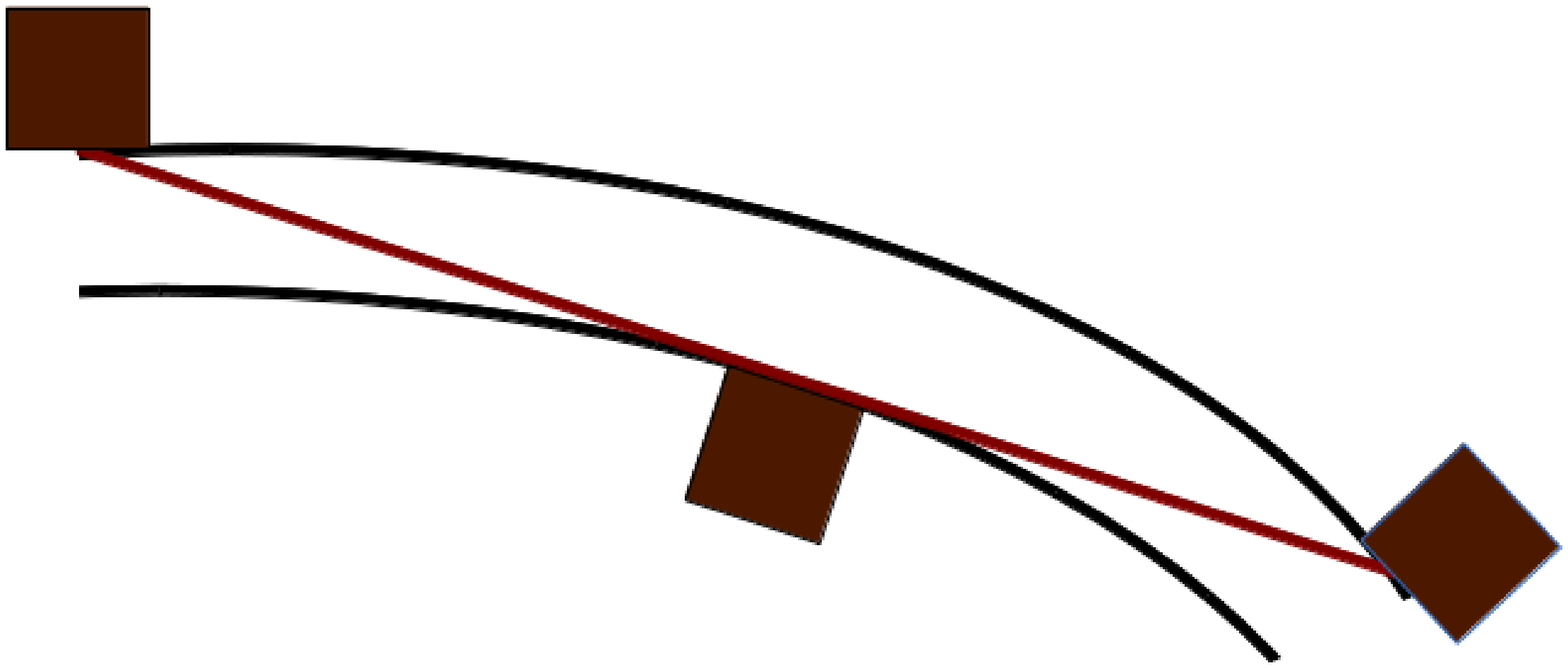}}
\resizebox{\columnwidth}{!}{%
\begin{tabular}{||c|c|c|c|c|c|c|c|c|c|c|c|c||c|c||}
\hhline{===============}
div   & $w_1$ & $w_2$  &  $w_3$ &  $w_4$ & $w_5$ & $w_6$ & $L_1$ & $L_2$ & $L_3$ & $L_4$ & $L_5$ & $L_6$ & $R_1$ & $R_2$ \\
      &  [cm] & [cm] & [cm] & [cm] & [cm] & [cm] & [m] & [m] & [m] & [m]  & [m] & [m]  & [m] & [m] \\
\hhline{===============}
$\pm$1.0\DEG & 1.9  & 4.0 & 4.0 & 4.0 & 4.0 & 1.4 & 4.2 & 17.3 & -    & -   & 15.2 & 6.3 & 1378 & 1299 \\
$\pm$0.5\DEG & 1.4  & 2.2 & 2.1 & 1.9 & 2.0 & 1.3 & 2.0 & 10.0 & 9.5  & 4.3 & 15.2 & 2.0 &  851 & 1287 \\
\hhline{===============}
\end{tabular}
}
\caption{(a) Schematic drawing of double-ballistic curved guide and
  optimization results for avoiding LoS at 20\,m and 45\,m
  (table). (b) Schematic drawing of line-of-sight calculation and
  metal block placement.}
\label{f_DoubleBallist_skizze}
\end{figure}

Particle swarm optimization has been used to optimize the guide
parameters shown in figure~\ref{f_DoubleBallist_skizze}. In order to
reduce the number of free parameters, the optimization started at
6.25\,m with a 2$\times$2\,cm\sq\,virtual source and horizontal and
vertical shape were optimized separately.  Only the horizontal shape
is scatched in figure~\ref{f_DoubleBallist_skizze}, the vertical shape
either matches the ballistic shape of the horizontal plane (without
the curvature) or follows a simple ellipse.  The first 4\,m of
horizontal guide (feeder) which focus into the first, 2\,cm aperture
were optimized separately.  Since the figure-of-merit in particle
swarm optimization maximizes either the neutron throughput or the
signal to noise ratio, but does not take beam quality measures like a
homogeneous phase space into account, PSO was used in combination with
parameter scans to obtain a high brilliance transfer as well as an
approximately smooth divergence distribution.

The performance is evaluated by the shape of the divergence
distribution as well as by the brilliance transfer. The latter is
calculated as the neutron intensity in a desired divergence range
obtained on the sample, divided by the neutron intensity in the same
divergence range and spatial area at the source. The sample area is
1$\times$1\,cm\sq.  The divergence range of interest for many
instruments is assumed to be $\pm$0.5\DEG\,or $\pm$1\DEG, so we will
examine both values in this study.

\subsection{Large divergence solution ($\pm$1\DEG)}

\begin{figure}[!htb]
\centering
\subfigure[Brilliance transfer]{\includegraphics[width=0.9\linewidth]{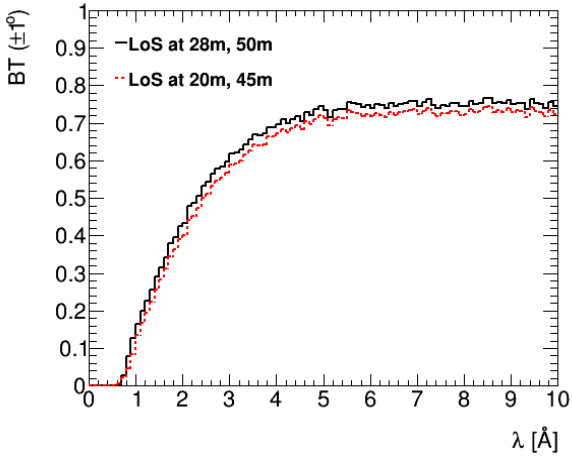}}
\subfigure[Horizontal divergence]{\includegraphics[width=0.9\linewidth]{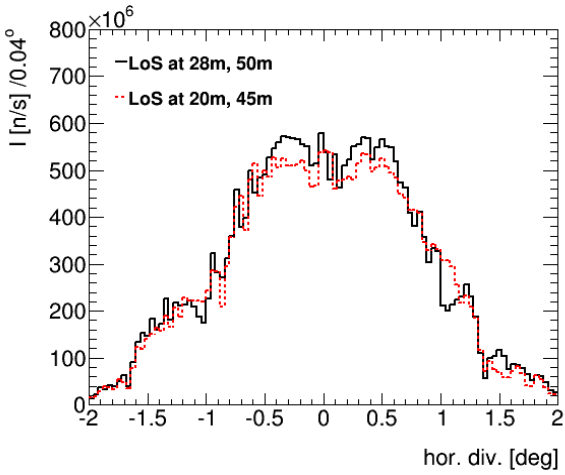}}
\caption{(a) Brilliance transfer (BT) and (b) horizontal divergence spectrum for the large divergence solution with different points of LoS avoidance as indicated in the figure legend.}
\label{f_BTdivy_1deg_LoSvgl}
\end{figure}

The best performance is obtained with the parameters given in the
table of figure~\ref{f_DoubleBallist_skizze}: a large divergence
requires a wide guide in order to minimize the number of reflections,
whilst at the same time the central focusing parts can be avoided if
the curved guide is of equal (or smaller) width as the central
aperture, i.e. 4\,cm in this case.  In addition, a larger fraction of
the guide can be curved if no central focusing is needed, entailing a
smaller curvature with larger radius.  The performance is shown in
figure~\ref{f_BTdivy_1deg_LoSvgl} for a square guide cross-section,
comparing one option that avoids LoS twice by the end of the guide at
49.75\,m with another that avoids LoS once in the common shielding
bunker at 20\,m and once more 5\,m before the sample position.  The
latter option is motivated by minimising background from secondary
particles produced in the shielding just before the sample
position. The former entails dividing the curved part up to 20\,m from
the source into two channels. These additional constraints are shown
to lead to only a minor loss of brilliance transfer, and the increased
curvature does not add any structure to the divergence spectrum. Hence,
the additional effort to avoid line of sight further away from the
sample position can and should be taken when designing an ESS
instrument of medium length.

\begin{figure}[!htb]
\centering
\subfigure[Brilliance transfer]{\includegraphics[width=0.9\linewidth]{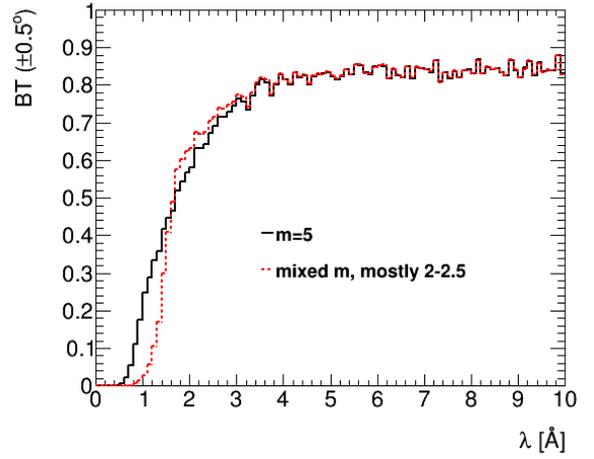}}
\subfigure[Horizontal divergence]{\includegraphics[width=0.9\linewidth]{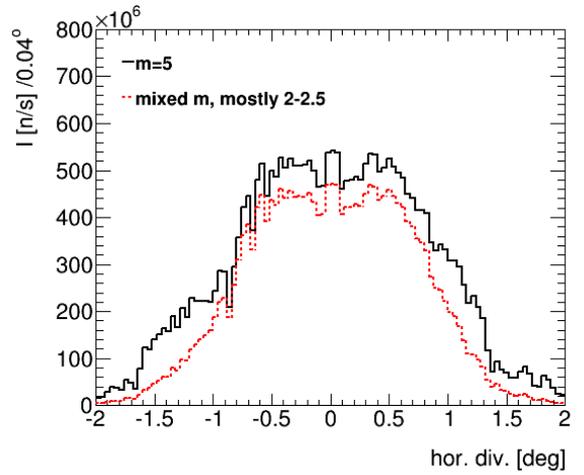}}
\caption{(a) Brilliance transfer (BT) and (b) horizontal divergence
  spectrum with different $m$-coating. Reduced coating tailored for
  2\,\Angst\,and $\pm$0.5\DEG\,divergence as example.}
\label{f_2A05deg}
\end{figure}

\paragraph{Supermirror coating}
For reasons of simplicity and to limit the parameter space, the
supermirror coating is fixed to a high value of $m$=5 in the
optimization of the guide shape. This is of course not necessary
throughout the whole surface area of the guide, in fact the
transmission of the guide can be tailored towards a specific
wavelength and divergence range by optimizing the coating accordingly:
this is demonstrated in figure~\ref{f_2A05deg}, which shows the BT and
divergence for the guide found above that avoids LoS at 20\,m and
45\,m. In the example shown here, the $m$ values are greatly reduced
to optimize the coating for 2\,\Angst\,neutrons in a limited
divergence range of $\pm$0.5\DEG. No BT is lost for $\lambda
\geq$2\,\Angst, with the $m$=5 coated area reduced to 1\% of the guide
surface and the major part of the guide coated wit $m$=2 (50\%) and
$m$=2.5 (25\%).

\subsection{Small divergence solution ($\pm$0.5\DEG)}
\begin{figure}[!htb]
\centering
\subfigure[Brilliance transfer\label{f_BTdivy_05degA}]{\includegraphics[width=0.9\linewidth]{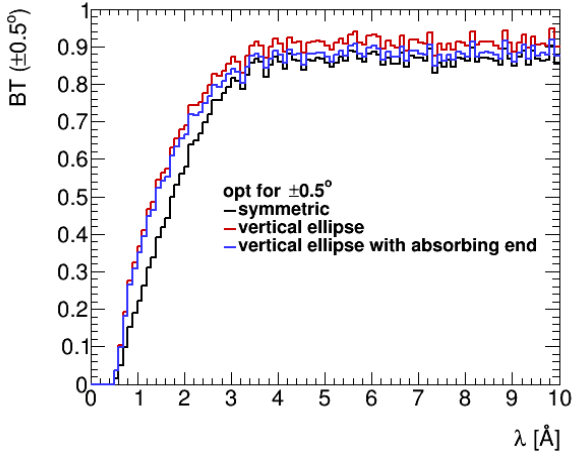}}
\subfigure[Vertical divergence\label{f_BTdivy_05degB}]{\includegraphics[width=0.9\linewidth]{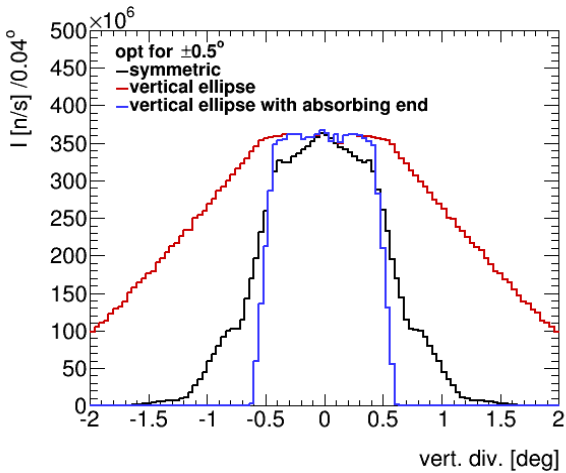}}
\caption{(a) Brilliance transfer (BT) and (b) vertical divergence
  spectrum for the small divergence solution with different vertical
  guide shapes: same height as width (black), vertical ellipse fully
  coated with reflecting surface (red), vertical ellipse with
  absorbing end sections (blue).}
\label{f_BTdivy_05deg}
\end{figure}

We now explore the effect of optimising the guide shape for a smaller
divergence range of $\pm$0.5\DEG\,.  The resulting guide parameters,
delivering the best performance, are shown in the second line of the
table in figure~\ref{f_DoubleBallist_skizze}.  The performance for a
quadratic guide cross-section is shown by the black line in
figure~\ref{f_BTdivy_05deg}. Note that for this solution, line of
sight is again avoided at 20\,m and 45\,m from the source.

Since the 2\,cm and 4\,cm wide apertures at 6.25\,m and 28\,m from the
source are motivated by choppers and have no obvious constraint in
vertical dimension\footnote{apart from an increase in price if the
  window height entails an increase of the chopper radius}, a
vertically elliptic shape without any height constraints can be
optimized as alternative to a symmetric guide.  As expected, in that
scenario the transmission efficiency increases slightly, shown by the
red line in figure~\ref{f_BTdivy_05degA}. However, the transmission of
background neutrons would be expected to markedly increase also, and
disproportionately to the thermal/cold neutrons of interest.  For
simplicity, we define signal neutrons as neutrons that hit the
sample within the desired phase space region, whilst background here
consists of neutrons that are transported to the sample plane outside
the desired divergence range (type I) or outside the sample area (type
II). Type I background can be seen in fig.~\ref{f_BTdivy_05degB}. As
shown by the blue line, this type of background can be effectively
suppressed by replacing the last vertical section of the focusing
guide with absorbing material (the last 2.85\,m in this example) since
the extra divergence the guide provides is not needed.

Type II background is also somewhat suppressed by this procedure, but
still larger with the vertical ellipse than with a symmetric guide:
the fraction of neutrons in the sample plain that hit the sample is
reduced from 0.54 (symmetric guide) to 0.46 (vertical ellipse with
absorbing ends). If no absorbing end section was used, this ratio
would even drop to 0.34.  In both of these cases one must also take
care about gamma background and scattering from the surface of the
absorber.

This example illustrates the importance of thinking about signal to
noise ratio rather than just maximizing the flux of neutrons on the
sample.

\subsection{Random misalignment and over-illumination}
Imperfections in the manufacturing and installation of neutron guides
can cause a reduced performance. Especially in long beamlines, such
minor effects can sum to significant impacts on brilliance transfer.
A previous study measured the effect of large misalignment in one
guide segment of a short miniature guide to extrapolate to the effect
of small misalignment in many segments of a longer guide
\cite{MisalignmentJapan1990}.  This approach has the advantage that
the misalignment effect in the miniature guide can be experimentally
verified, but the results for larger scale systems need to be
validated.  A simulation study of a long straight guide
\cite{LossMechanisms2001} concludes that misalignment deviations up to
0.02\,mm can be tolerated for a 40\,m long
150$\times$50\,mm\sq\,neutron guide, i.e. a relative misalignment of
only 0.04\%.  An assumed absolute misalignment of 50\,$\mu$m is
generally used as the standard installation specification by neutron
optics vendors, using laser trackers and/or theodolites.  This is
under laboratory conditions, and in the field we expect that thermal
expansion and floor loading cause larger deviations even on relatively
short timescales.  Nonetheless, in the example of the 2.2\,cm guide
width of the small angle solution from the previous section, this
50\,$\mu$m random misalignment constitutes a 0.2\% geometrical
uncertainty on a curved guide.  We therefore explore a strategy to
mitigate misalignment effects in the following section.

The concept of overilluminating subsequent guide sections to prevent
misalignment effects has been successfully employed at
JPARC~\cite{PrivCommArai}, and we examine it here for the case of a
double curved guide as described in the previous section.  A detailed description and a possible mitigation by overillumination is given in  \ref{A_sec_Misalignment}, here we give a summary:
\begin{enumerate}
\item The relative transmission of a poorly aligned guide compared to
  a perfectly aligned guide can be described by the function
  \begin{equation}
    T=\left( 1-\frac{2\delta}{w} \right)^{N} \cdot (1+N\frac{f}{100}\frac{\delta/50\mu m}{w}) \notag
  \end{equation}
  where the same horizontal and vertical misalignment $\delta w =
  \delta h =: \delta$ is assumed for a guide of width $w$ and height
  $h=w$ constructed from $N$ independent sections. The factor $f=0.20
  \pm 0.02$ is the relative difference between gaussian and fixed
  misalignment per guide segment, per 50\,$\mu m$ misalignment and per
  guide width (in cm), found by simulation (see
  appendix~\ref{A_sec_Misalignment}).
\item Overillumination of subsequent guide sections reduces the
  beamloss significantly for guides with small cross-sections, while
  large guides (around 10$\times$10\,cm\sq) do not benefit from this
  approach.
\item The beamloss caused by misalignment is not increased if the
  guide is curved. Losses from overillumination increase in curved
  guides.
\item The beamloss caused by misalignment of a ballistic guide is
  similar to (but slightly larger than) the beamloss in a straight
  guide with a guide cross-section of $(w_{min}+w_{max})/2$.
\item The beamloss caused by misalignment of a multi-channel guide is
  slightly higher than expected from the cross-section of a single
  channel, an additional loss due to the overlap of channel separating
  blades is visible.
\item Spatial misalignment significantly dominates angular misalignment.  More
  specifically, the effect of an angular misalignment corresponding to
  a spatial misalignment per segment length,
  $\delta\alpha=\delta/L_{segment}$, is negligible compared to the
  spatial misalignment. This is also true for curved guides.
\end{enumerate}

Applying these general findings to the guide of the previous section,
which is double-curved in the horizontal direction with a maximum
guide width of about 2\,cm, and vertically an ellipse with a maximal
guide height of 12\,cm and a minimal one of about 9\,cm (apart from
the absorbing end section), it follows that an overillumination
approach would not be beneficial in the vertical direction due to the
large height and hence small relative misalignment of the guide,
considering an alignment precision of 50\,$\mu m$. In contrast,
overillumination is promising in the horizontal direction due to the
comparably small guide width.

However, the largest misalignment is expected to be in the vertical
direction, as the dominant process is settling of the ground under
heavy shielding and construction loads.  The horizontal movement is a
second order effect caused by pivoting or rotations about a support
point.

\begin{figure}[!htb] 
\centering
\subfigure[Loss from vertical (v) and horizontal (h) misalignment.\label{f_MisalignmentB}]{\includegraphics[width=0.9\linewidth]{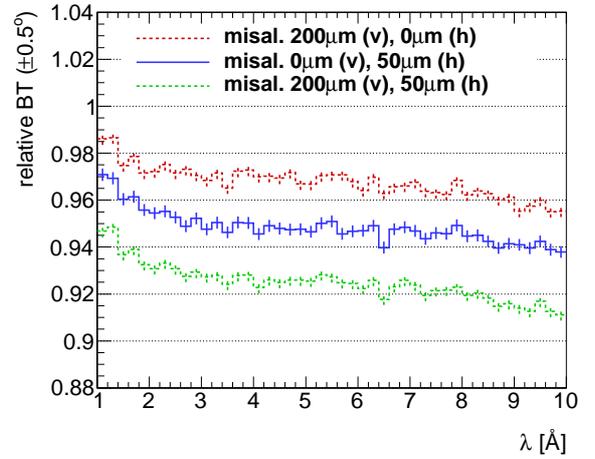}}
\subfigure[Mitigation by and loss from overillumination approach. \label{f_MisalignmentC}]{\includegraphics[width=0.9\linewidth]{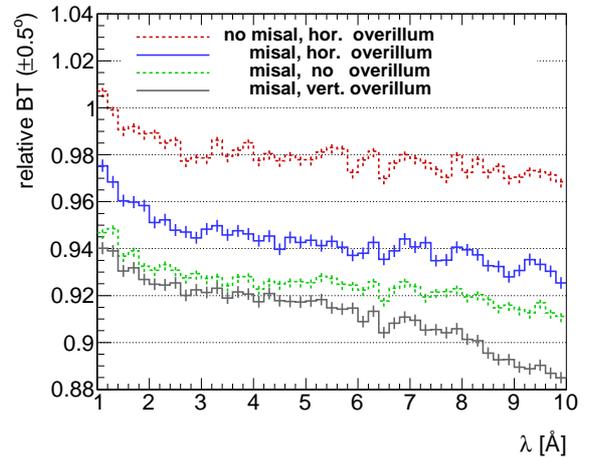}}
\caption{Impact of misalignment and mitigation by overillumination: (a) Loss from vertical (v) and horizontal (h) misalignment. (b) Mitigation by and loss from overillumination approach.}
\label{f_MisalignmentApplication}
\end{figure}

At ESS, an elastic vertical ground movement of 3\,mm is anticipated
when loading the experimental floor with shielding.  In addition, a
further 3\,mm of creep is expected over 10 years.  If we assume the
possibility for re-alignment of a guide fits the facility schedule
every 6 months, the latter results in additional 150\,$\mu
m$. Therefore in the following, 200\,$\mu m$ are assumed vertically
and 50\,$\mu m$ horizontally. In the simulation, guide pieces are
shifted against each other by random offsets calculated from a
gaussian probability distribution with these misalignment values as
standard deviation, centered around 0\,$\mu m$. The guide is cut into
2\,m long sections. The impact of misalignment ist shown in
figure~\ref{f_MisalignmentB} as the ratio of the brilliance transfer
on the sample with and without misalignment. As expected from the
general study, the horizontal misalignment of 50\,$\mu m$ causes a
larger beamloss than the vertical misalignment of 200\,$\mu m$ due to
the much smaller guide width than height. Both together lead to a
brilliance loss of 5\% to 9\% between 1\,\Angst\,and 10\,\Angst.
Taking the horizontally ballistic shape into account, the transmission
expected from the formula above with 50\,$\mu m$ misalignment in the
horizontal direction alone is about 96\% of the transmission of the
perfectly aligned guide. This is also seen in the simulation as the
mean transmission ratio; when however the brilliance transfer in
$\pm$0.5\DEG\,is compared instead, the loss due to misalignment is
somewhat larger because in a curved guide, predominantly Garland
reflected neutrons with small divergence are affected by misalignment.

A possible mitigation by (horizontal) overillumination is shown in
figure~\ref{f_MisalignmentC}: The horizontal overillumination itself
causes a beamloss between 2\% and 3\% depending on the wavelength. If
the guide pieces are poorly aligned, the 5\%-9\% loss from before is
reduced to 3\%-7\% , i.e. only a 2\% gain is opposed to a 2\%-3\%
possible loss if the guide is better aligned than expected.

The effect of a vertical overillumination is also found, as expected
from the general study, to have a negative impact on transmission if
it is over-used.  The reason for these counter-intuitive effects is
the beam extraction efficiency as the guide entrance increases in
size.  Consequently, we anticipate that overillumination should be
used sparingly, at one or two key locations where large floor
movements might be anticipated (e.g.  joins in the floor foundations).

\section{Double line-of-sight with short instruments (20\,m)}
\label{sec_Generic20m}

In this section, we examine some options for short beamlines with a
sample position at 20\,m distance from the source.  It is of interest
for all instruments to lose line of sight as close to the source as
possible.  This both minimises background and takes advantage of the
common shielding in the bunker, to minimise instrument costs.

The options we study all lose line of sight twice, and have a monolith
insert comprised of a simple, 4\,m long guide of constant cross
section, starting 2\,m from the source.  In all cases, this means that
any curved or inclined guide sections are at least 6\,m from the
source, outside the monolith. The guide width is fixed to 3\,cm.
Unless stated otherwise, the supermirror coating is fixed at $m=$\,2
in straight guide parts and $m=$\,3 in curved parts, again to minimise the cost of the
system.  Five design options are compared and evaluated at the same
point, i. e., at the end of the common shielding bunker.

\begin{figure}[!htb] 
\centering
\subfigure[\label{f_20mSchema}]{\includegraphics[width=0.7\linewidth]{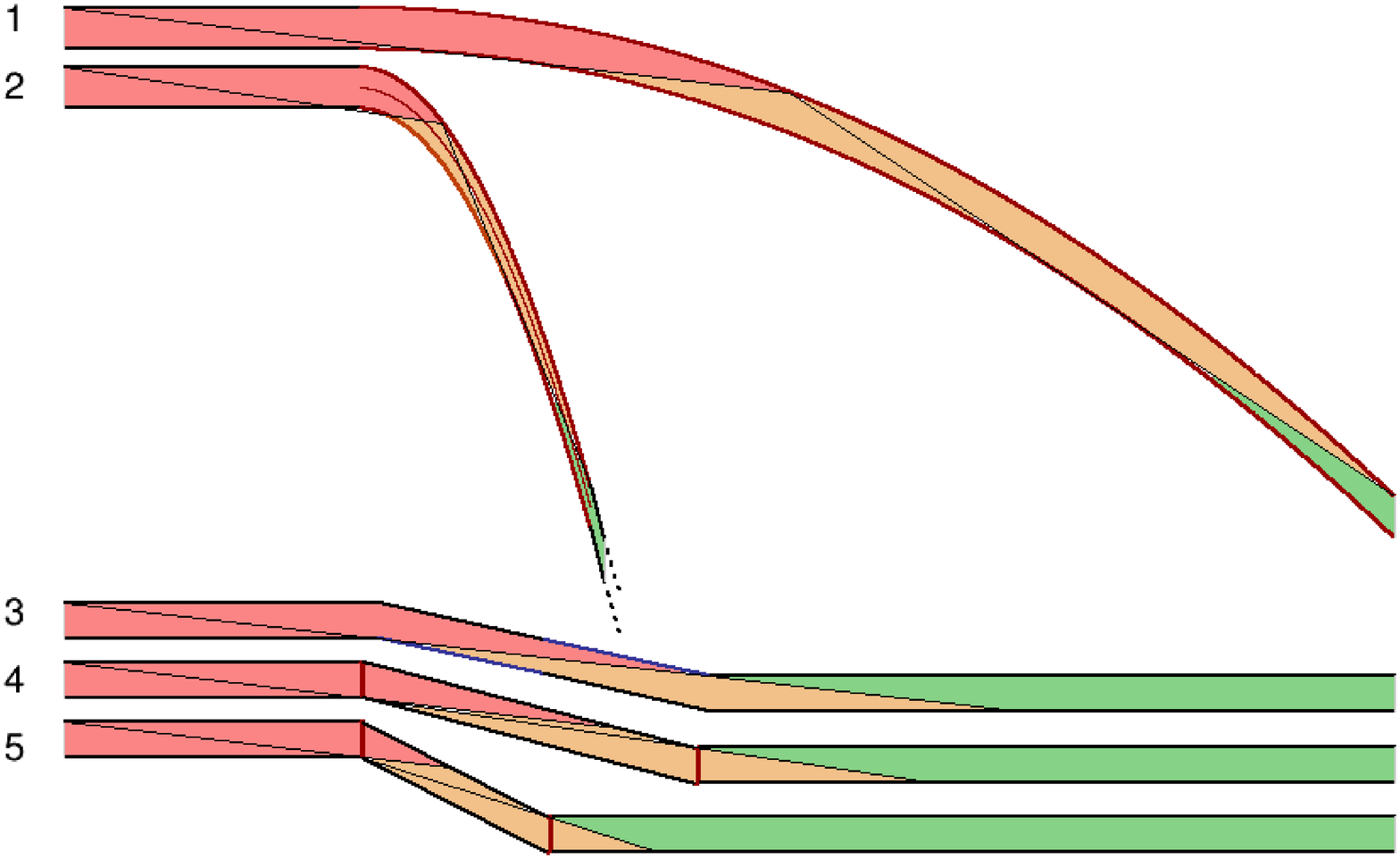}}
\subfigure[\label{f_DivSolidBender}]{\includegraphics[width=0.25\linewidth]{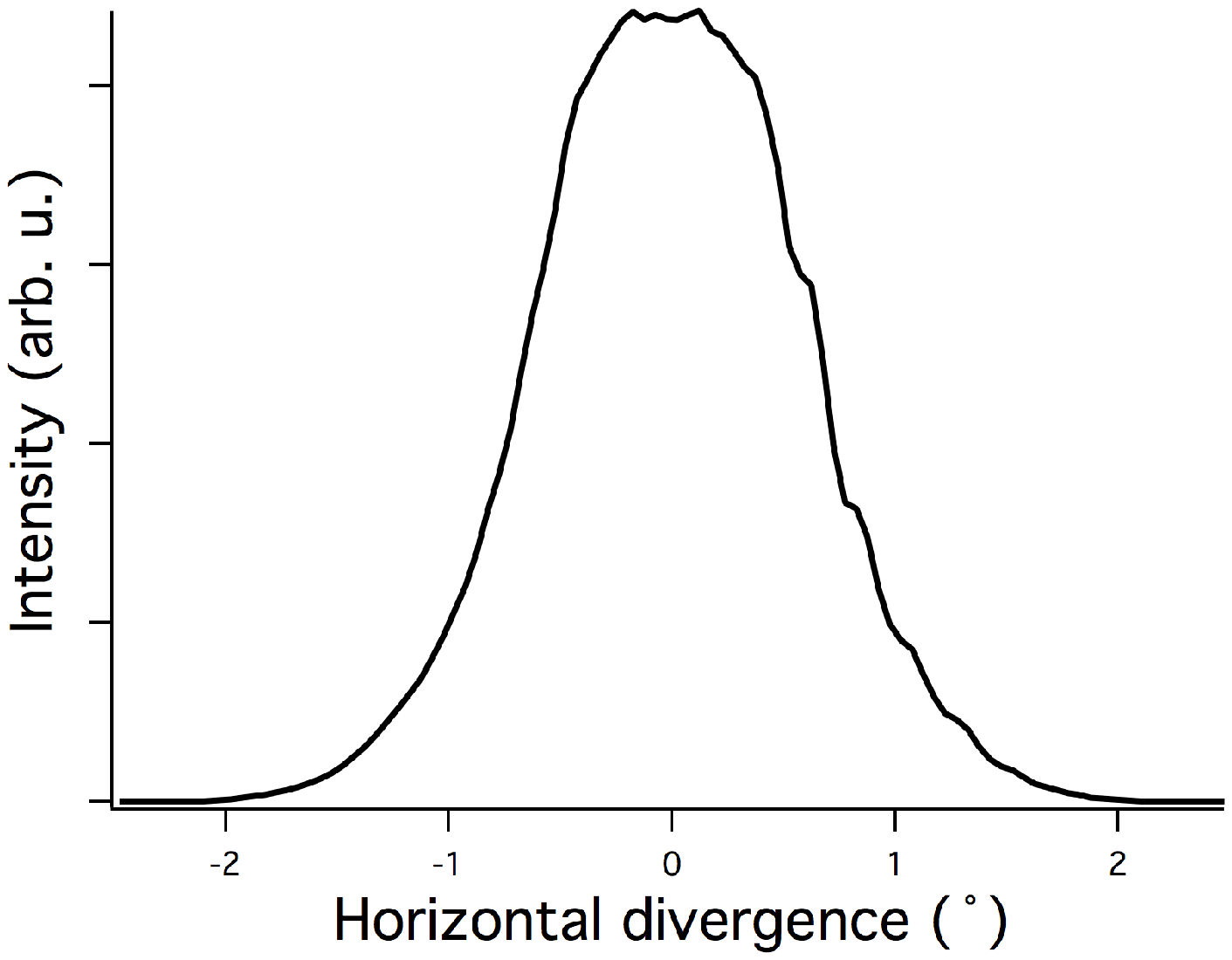}}
\caption{(a) Conceptual illustration of systems 1 - 5 (top to bottom). Red regions are within direct line of sight, orange regions have lost line of sight once, and green regions twice.  Red guide walls correspond to curved sections. \newline (b) Horizontal divergence distribution directly after the first bender of system 4.
}
\end{figure}

The options considered are as follows:
\begin{description}
\item[system 1] consists of a simple curved guide.  This system only
  just satisfies the twice line of sight condition, losing 2nd line of
  sight right at the exit point, and having a radius of curvature of
  239.26\,m.  No instrument should build this, but we study this as an
  upper limit to performance with the most favourable curvature for
  transport efficiency.
  
\item[system 2] is a single multi-channel bender, followed by a
  straight guide. The bender is designed such that the critical length
  needed to go out of line of sight is 2\,m. This results in a
  curvature radius of 14.2\,m for a 3.12\,m long bender with 12
  channels.
  
\item[system 3] includes a double-bounce mirror in form of a kinked
  guide, with mirror surface and absorbing sections, such that the
  length of the inclined guide section is twice the length of the
  mirror section. Thus the inclined section starts at 6.28\,m from the
  source and ist 2$\times$2.23\,m long, the inclination angle is
  0.79\DEG\,and the coating of the mirror $m=$\,4. Absorbing sections
  are designed to suppress large divergence neutrons.
  
\item[system 4] uses two solid state benders of 5\,cm length with 300
  channels of 100\,$\mu$m thickness, m=3 between channels, and a
  radius of 3.125\,m. The two benders are similar in concept to the
  work of Krist \emph{et al} \cite{KRIST-SOLID-STATE-BENDERS}.  They
  are separated by a straight guide of 4.5\,m length. The idea of the
  bender design is to adjust the curvature such that it matches the
  critical one for the channels to avoid gaps in the transmitted
  divergence, which is verified in figure~\ref{f_DivSolidBender}. The
  channel walls are made of silicon, simulated options for substrate
  materials inside the channels are vacuum (ideal maximum
  transmission), silicon, and carbon.
  
\item[system 5] is similar to system 4, but with only 150 bender
  channels of 200\,$\mu$m thickness coated with m=3, a distance
  between benders of 2.5\,m and bender radii of 1.562\,m.
\end{description}

Figure~\ref{f_20mSchema} gives a schematic overview over these systems
and the regions in which the guide is out of line of sight twice
(green), once (orange), and compared to a straight beamline
(red). Curved guide sections are marked by red lines, so the very
short solid benders of systems 4 and 5 appear as red stripes in an
otherwise z-like geometry. For better visibility, not all bender
channels are shown for system 2 and the guide is cut as it gets out of
line of sight a second time, but the large offset of the beamline
compared to the other options is still obvious.

Note that for all systems involving curved sections, i.e. all but
system 3, the curvature and loss of line of sight points have been
calculated with an increased guide width of 3.5\,cm instead of the
used (simulated) 3.0\,cm in order to force all neutron trajectories to
pass a minimum amount of shielding material. In
figure~\ref{f_20mSchema}, an increased width is also drawn in systems
1 and 2 to illustrate the theoretical regions, while systems 4 and 5
illustrate the extra margin by leaving the ``red'' regions well before
the second kink, as done in the simulations. System 3 is the only one
which is both drawn and simulated without any extra margin, so an
increased background can be expected there.

In case of multi-channel benders, cross-talk between channels has been
modelled in the simulation.

\begin{figure}[htb!] 
\centering
\subfigure[\label{f_20mDR_s45}]{\includegraphics[width=0.9\linewidth]{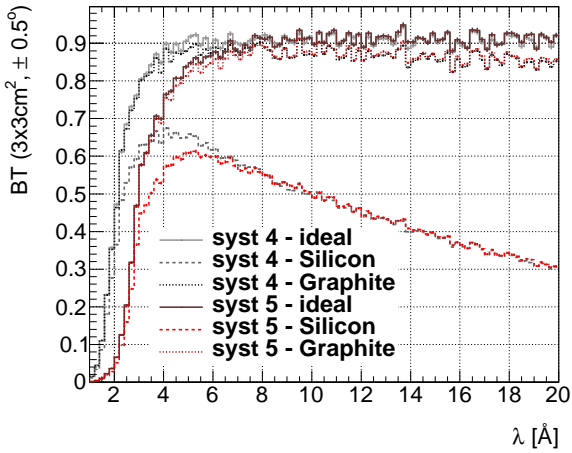}}
\subfigure[\label{f_20mDR_all}]{\includegraphics[width=0.9\linewidth]{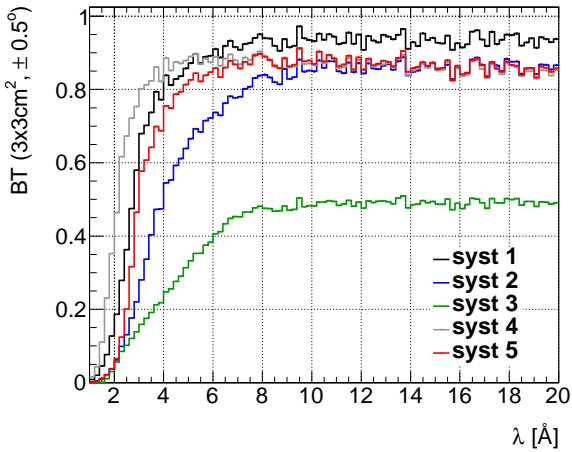}}
\caption{(a) BT of solid benders including the absorption in vacuum, silicon or graphite. (b) Comparison of the BT of the 5 systems.}
\label{f_20mDR}
\end{figure}

Figure~\ref{f_20mDR_s45} compares different options for the solid
benders of systems 4 and 5: the ideal case with vacuum between the
channels (solid lines) gives an excellent performance, but including
the absorption of silicon (coarse dotted lines) reduces the brilliance
transfer dramatically. Therefore, it is worth considering alternative
materials. The third option in figure~\ref{f_20mDR_s45} shows the BT
with solid benders made of graphite, which has a much lower absorption
cross-section than silicon and therefore a very high brilliance
transfer.  Indeed, graphite matches the performance almost of the
perfect (vacuum) case.  Construction of a graphite solid state bender
will depend on the possibility to manufacture graphite wafers with
sheets of constant thickness and a small enough surface roughness, and
it remains to be seen whether the SANS from such a system is
tolerable\footnote{The feasibility is currently under investigation
  with a prototype study.}.

The brilliance transfer of all five systems, evaluated in the whole
3$\times$3\,cm\sq\,guide exit region for a divergence of $\pm$0.5\DEG,
is shown in figure~\ref{f_20mDR_all}. The highest brilliance transfer
is seen for system 1, as expected.  Almost the same performance could
be achieved with the graphite solid bender systems, losing only about
10\% more cold neutrons and (possibly) fewer thermal neutrons than the
simple curved guide, even though line of sight is avoided much closer
to the source. The multi-channel bender delivers equivalent
performance for cold neutrons only, while losing a significant amount
of thermal neutrons. The double-bounce mirror system is seen to cause
unacceptably high transmission losses in the whole wavelength band.

Since all these systems lose line of sight at different distances from
the source, it is instructive to compare them to simple curved neutron
guides. The curved section is split into two pieces in order to
compare the effect of an s-shape and c--shape curved
guide. Table~\ref{t_20mSystems} summarises the points of line of sight
closure as well as bender properties needed to avoid line of sight at
the same distances from the source. As in systems 1-5, the first 4\,m
are just a constant guide, so the first bender extends between 6\,m
and the first point of line of sight, and the second bender between
the first and second point of line of sight. The coating is $m=$\,3 on
curved surfaces and $m=$\,2 otherwise. The number of channels is
chosen such that the curved guide is as similar as possible to the
original system, i.e. one channel in systems 1 and 3, and 12 channels
in system 2. For comparison with the solid benders of systems 4 and 5,
the number of channels is chosen such that the same $\lambda_c$ is
achieved.

\begin{table}[htb!]
\centering
\resizebox{\columnwidth}{!}{
\begin{tabular}{||c||c|c||c|c|c|c|c|c|c||}
\hhline{==========}
System  & 1$^\mathrm{{st}}$ LoS & 2$^\mathrm{{nd}}$ LoS  &  $L_{B1}$ & $R_{B1}$ & $N_{B1}$  & $L_{B2}$ & $R_{B2}$ & $N_{B2}$ & $\lambda_c$\\
\hhline{==========}
1 & 11.81\,m & 20.00\,m  & 5.8\,m & 238.0\,m & 1 & 8.2\,m & 239.5\,m & 1 & 3.0\,\Angst \\
2 &  7.12\,m &  9.12\,m  & 1.1\,m & 14.2\,m & 12 & 2.0\,m & 14.2\,m & 12 & 3.6\,\Angst \\
3 &  9.74\,m & 14.97\,m  & 3.7\,m & 114.0\,m & 1 & 5.2\,m & 98.0\,m & 1  & 2.0\,\Angst \\
4 &  9.53\,m & 14.42\,m  & 3.5\,m & 104.0\,m & 9 & 4.6\,m & 75.2\,m & 12 & 1.5\,\Angst \\
5 &  7.22\,m & 11.49\,m  & 1.2\,m & 16.5\,m & 14 & 4.3\,m & 65.1\,m & 4  & 3.1\,\Angst \\
\hhline{==========}
\end{tabular}
}
\caption{1$^\mathrm{{st}}$ and 2$^\mathrm{{nd}}$ point of line of sight avoidance as well as lengths ($L_{B1,2}$), radii ($R_{B1,2}$) and number of channels ($N_{B1,2}$) of corresponding benders $B1$ and $B2$. }
\label{t_20mSystems}
\end{table}

\begin{figure}[!htb] 
\centering
\subfigure[\label{f_20m_CS}]{\includegraphics[width=0.9\linewidth]{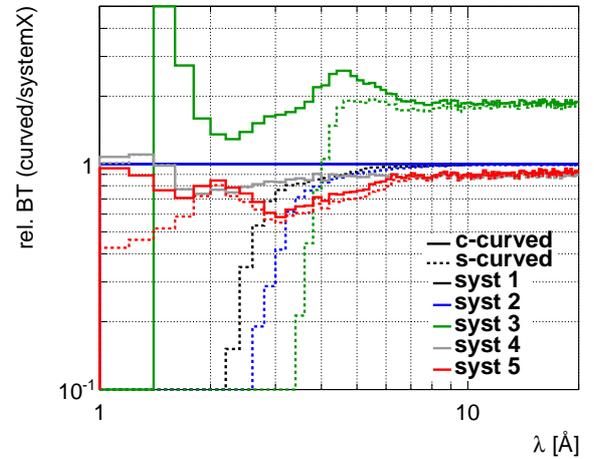}}
\subfigure[\label{f_Syst4}]{\includegraphics[width=0.9\linewidth]{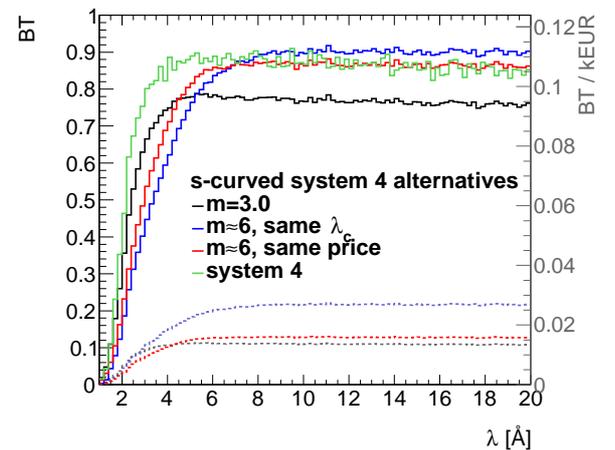}}
\caption{(a) Relative BT of curved guides w.r.t. systems 1-5. C-curved lines of systems 1 and 2 lie on top of each other. (b) BT of system 4 alternative s-benders with different coating. Light coloured dotted lines show BT per k\euro\,(y-axis on right-hand side).}
\end{figure}

Figure~\ref{f_20m_CS} shows the ratio of the brilliance transfer (BT) obtained with an s- or c-curved guide to the BT of the corresponding system 1-5. Since systems 1 (black) and 2 (blue) are c-curved options themselves, the c-ratio is just a constant line at 1, and the dotted lines directly show the transmission loss in short wavelengths of an s-bender compared to a c-bender. The s-shape reaches a BT of 90\% of that of the c-shape only for wavelengths longer than 4.5\,\Angst.  This is the expected behaviour, since s-benders have the advantage of sharper short-wavelength rejection. 

The kinked guide (system 3, green lines) is the only one which is a
worse solution than a simple curved guide, but for
$\lambda<$\,4\,\Angst\,it performs better than a corresponding
s-bender. Note that the sharp cut-off at 1.5\,\Angst\,is caused by
zero statistics in the double-bounced mirror brilliance transfer, not
the BT of the curved guide\footnote{Infinity is set to zero.}.

The situation is slightly different for the short solid benders
(systems 4 and 5), which are again evaluated assuming absorption in
carbon: a simple c-shaped bender gives 90\% of the brilliance transfer
of those systems in most of the wavelength band (system 4) or for
$\lambda>$\,6\,\Angst\,(system 5), and a minimum BT ratio of 75\%
(system 4) to 60\% (system 5). Hence the solid benders perform better
for thermal neutrons, while the more classical systems are a good
alternative for cold neutrons without the necessity of further R\&D to
find the best suitable material in terms of unwanted absorption and
scattering in the bender material. Interestingly, the corresponding
s-bender delivers the same BT as the c-shaped guide in case of system
4, therefore the solid and dotted line lie on top of each other in most
of the wavelength band.

\subsection{Cost considerations} 

The supermirror coating of $m=$\,2 in straight and $m=$\,3 in curved
guide sections was chosen to keep the cost low, in case of
multi-channel benders it can however be cheaper to use a higher
coating with fewer channels. Figure~\ref{f_Syst4} shows the example of
system 4, the BT of which is compared to s-benders with the same
$\lambda_c$ and either $m=$\,3 or $m=$\,6. The $m=$\,6 option costs
only about 60\% of the $m=$\,3 option (considering the price of the
benders alone) and gives a higher brilliance transfer for long
wavelengths, due to having fewer channels. The short wavelength
transmission is, however, reduced because of the lower reflectivity at
$\lambda_c$. The light coloured, dotted lines show the brilliance
transfer per k\euro\, is highest for the $m=$\,6 option.

In comparison, an $m=$\,6 option costing about the same as the $m=$\,3
version is shown to give the same brilliance transfer as $m=$\,3 for
long wavelengths, but the brilliance transfer decreases considerably
at short wavelengths.

\section{Conclusions}

Generic neutron guides for ESS instruments of short and medium length
were optimized for good performance at low cost under the condition
that a direct line of sight to the source is avoided twice. These
guides can serve as a baseline and are used to develop concepts
addressing common neutron optics challenges at ESS.

It was shown that medium length
instruments can avoid line of sight once within the bunker to use the
common shielding as well as once more well before the sample position
to avoid background without much loss of brilliance transfer. It was
further illustrated that the supermirror coating should be tailored
for the wavelength and divergence in mind, to effectively save money and
suppress short wavelength neutrons while retaining useful
neutrons, and that the coating and guide shape should be optimized
such that not only the number of useful neutrons on the sample is
maximized but rather the signal over background is
considered.

We have shown that transmission losses from random misalignment can to
some extent be mitigated by overillumination, but only for a small
guide cross-section where the relative misalignment is
greatest. Otherwise, the losses caused by overillumination, through
poor beam extraction efficiency from having too large a guide
entrance, can cause comparable or even larger losses than the
misalignment if applied between all sections. Therefore the strategy
for ESS is to identify high risk positions in the guide and only apply
overillumimation there.  This means that to achieve alignment over
large distances, it is likely that ESS will need to explore further
options in the near future.

For short instruments (or those with very low background tolerance),
the line of sight to the source can be avoided twice within 20\,m from
the source. A good brilliance transfer can be achieved for
2\,\Angst\,and longer wavelengths without need for high supermirror
coating. Different approaches for leaving line of sight as quickly as
possible are presented and show that double-curved beamlines perform
much better than double-kinked, and that short solid multi-channel
benders are a promising approach provided that the absorption in the
bender material can be controlled. To this end, a prototype study
involving carbon substrates has been started.

\section{Acknowledgments}

We would like to thank Swiss Neutronics, Mirrortron and S-DH for their direct and indirect input into various aspects of the ESS development work leading up to this study.

\label{Bibliography}
\bibliographystyle{unsrtnat}
\bibliography{References_preprint}
\begin{appendix}
\input{Results_GeneralMisalignmentStudyMerge_preprint} 
\end{appendix}
\end{document}

%% file: Results_GeneralMisalignmentStudyMerge_preprint.tex
\section{Random Misalignment}\label{A_sec_Misalignment}

The impact of misalignment on guide performance is simulated for a
generic 20\,m long guide composed of 1\,m long segments. The simulated
neutron source is the 12\,cm high TDR moderator, if not explicitly
stated otherwise. This moderator is chosen to ensure a good
illumination of the guide entry up to guide cross-sections of
10$\times$10\,cm\sq. The results are presented in the form of the
relative transmission of the guide, including misalignment between
guide segments, compared to the same guide geometry with no
misalignment. As this is a comparative study, the absolute brilliance
of the source can be ignored.

\subsection{Spatial misalignment}

Spatial misalignment is simulated as independent simultaneous shifts
in horizontal and vertical direction. The magnitude of the shift is
determined by a gaussian shaped random number with mean value 0 and
standard deviation $misalignment$.

\subsubsection{Straight Guides: Comparison of misalignment function with VITESS simulations}

The transmission $T'$ of a constant guide with misalignment, compared
to the transmission $T_{0}$ of the same guide without misalignment,
can be described by the following
function~\cite{I_ANDERSON_MISALIGNMENT_FUNCTION}:
\begin{equation}\label{eq_fctn1}
T=T'/T_{0}=\left( 1-\frac{\delta w}{w}-\frac{\delta h}{h} \right)^{N}
\end{equation}
where $N$ is the number of guide pieces, $w$ and $h$ are guide width
and height and $\delta w$ and $\delta h$ the horizontal and vertical
misalignment. In the following, $\delta w = \delta h =
\delta$. \newline

\begin{figure}[!htb]
\subfigure[20\,m guide\label{f_Theory1a}]{\includegraphics[width=0.9\linewidth]{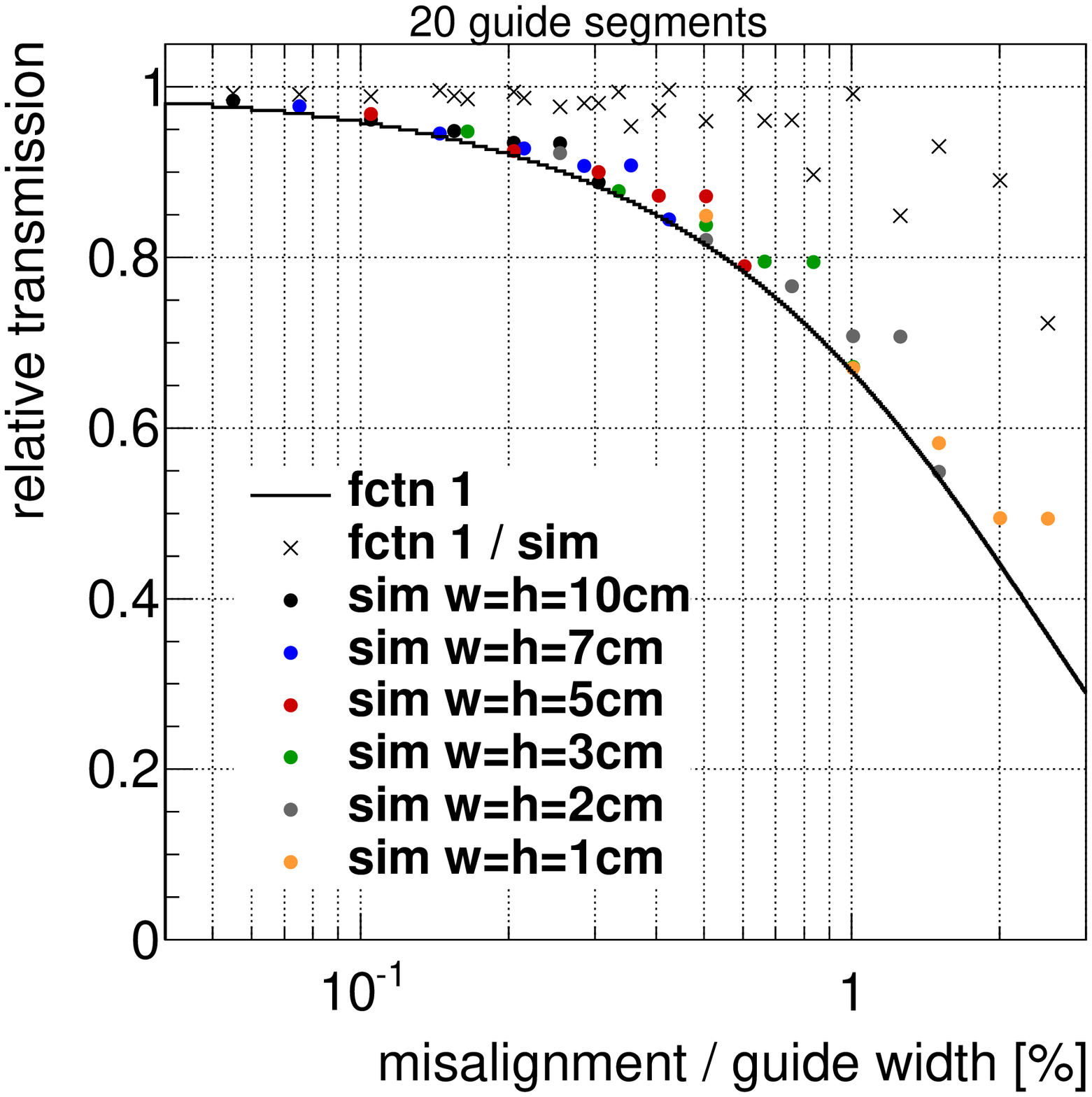}}
\subfigure[150\,m guide\label{f_Theory1c}]{\includegraphics[width=0.9\linewidth]{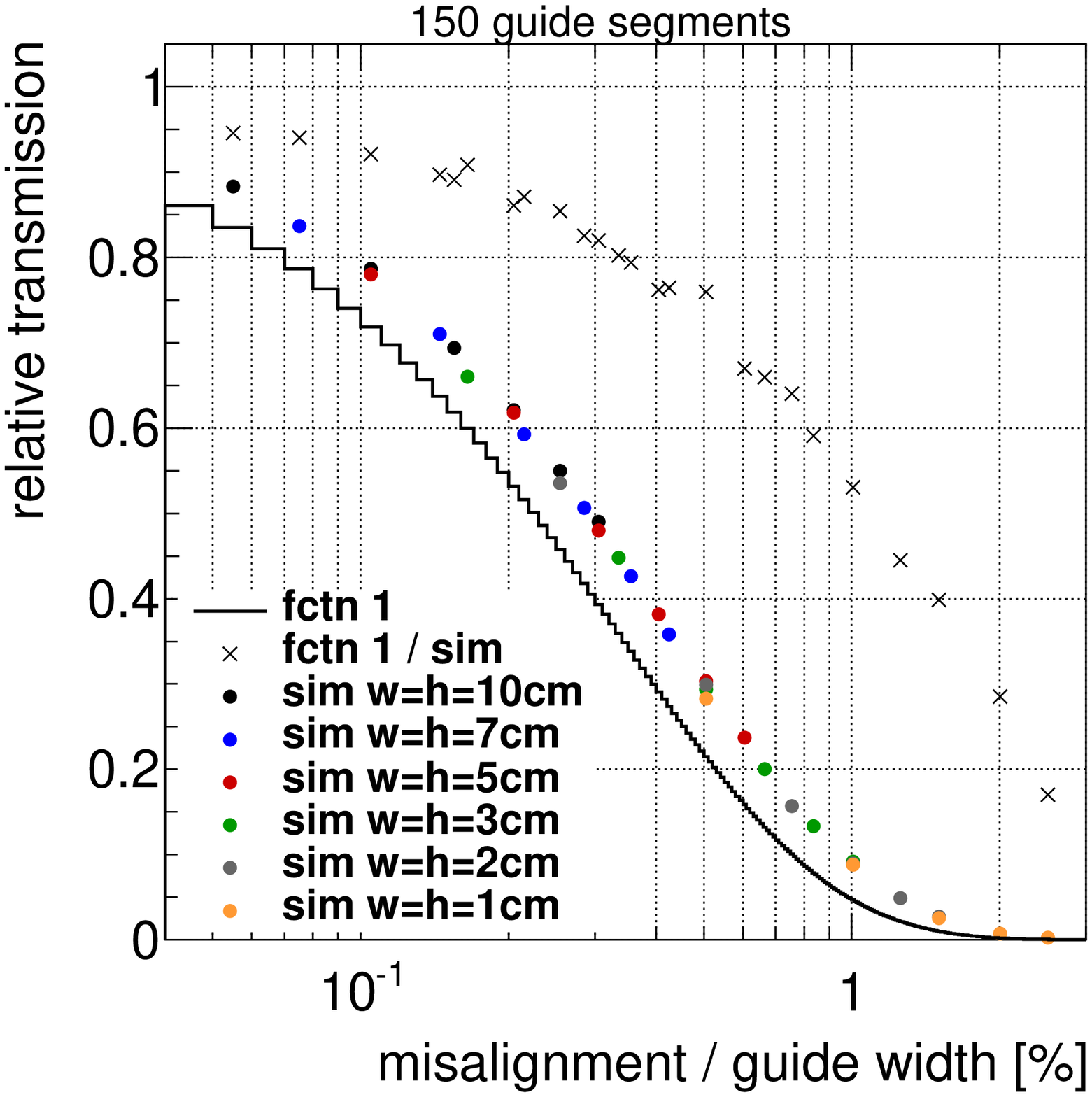}}
\caption{Relative transmission of a (a) 20\,m and a (b) 150\,m guide
  with misalignment compared to the transmission calculated with
  function~\ref{eq_fctn1} (solid lines), showing that the simple model
  applies to short guides but does not accurately predict the
  behaviour of long ESS guides so well.  Round markers are the
  simulation results, and crossed markers are the ratio, with colours
  representing different guide dimensions as shown in the legend.}
\label{f_Theory1}
\end{figure}

Figure~\ref{f_Theory1a} shows the transmission for a 20\,m long guide
with different cross-sections built out of 1\,m long segments. Since
the transmission loss depends only on the amount of misalignment
relative to the guide width, the relative transmission is shown as a
function of percental misalignment. The transmission predicted by
function (1) is within 10\% of the simulation for less than 1\% of
misalignment in a 20 piece guide, but systematically below the
simulated transmission. The discrepancy increases with the amount of
misalignment and with the guide length, as seen in
figure~\ref{f_Theory1c}. Hence function~\ref{eq_fctn1} is only an
adequate estimate of the maximum loss expected for short guides made
from a small number of segments.

\begin{figure}[!htb]
\subfigure[20\,m guide, fct 2]{\includegraphics[width=0.9\linewidth]{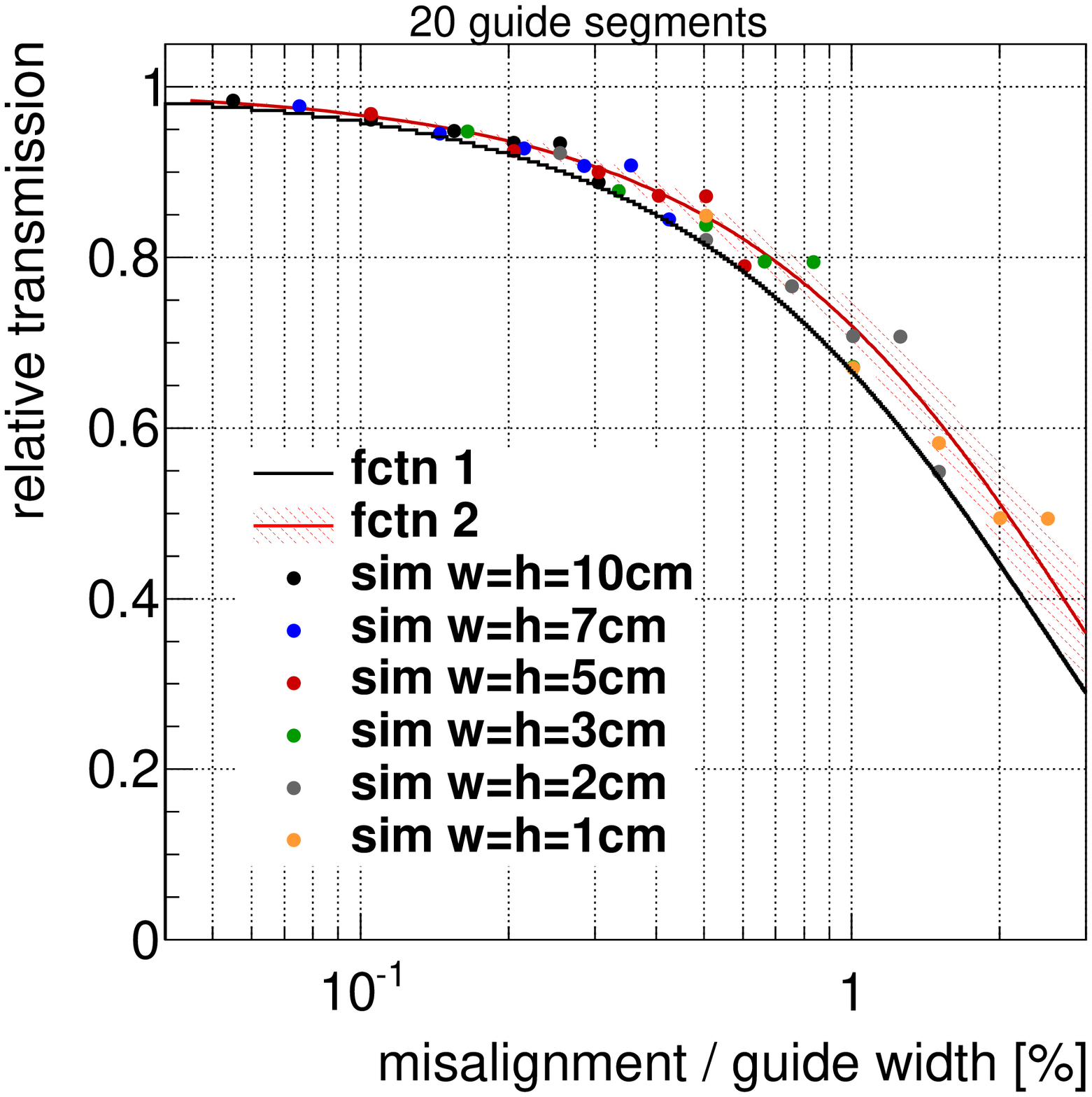}}
\subfigure[150\,m guide, fct 2]{\includegraphics[width=0.9\linewidth]{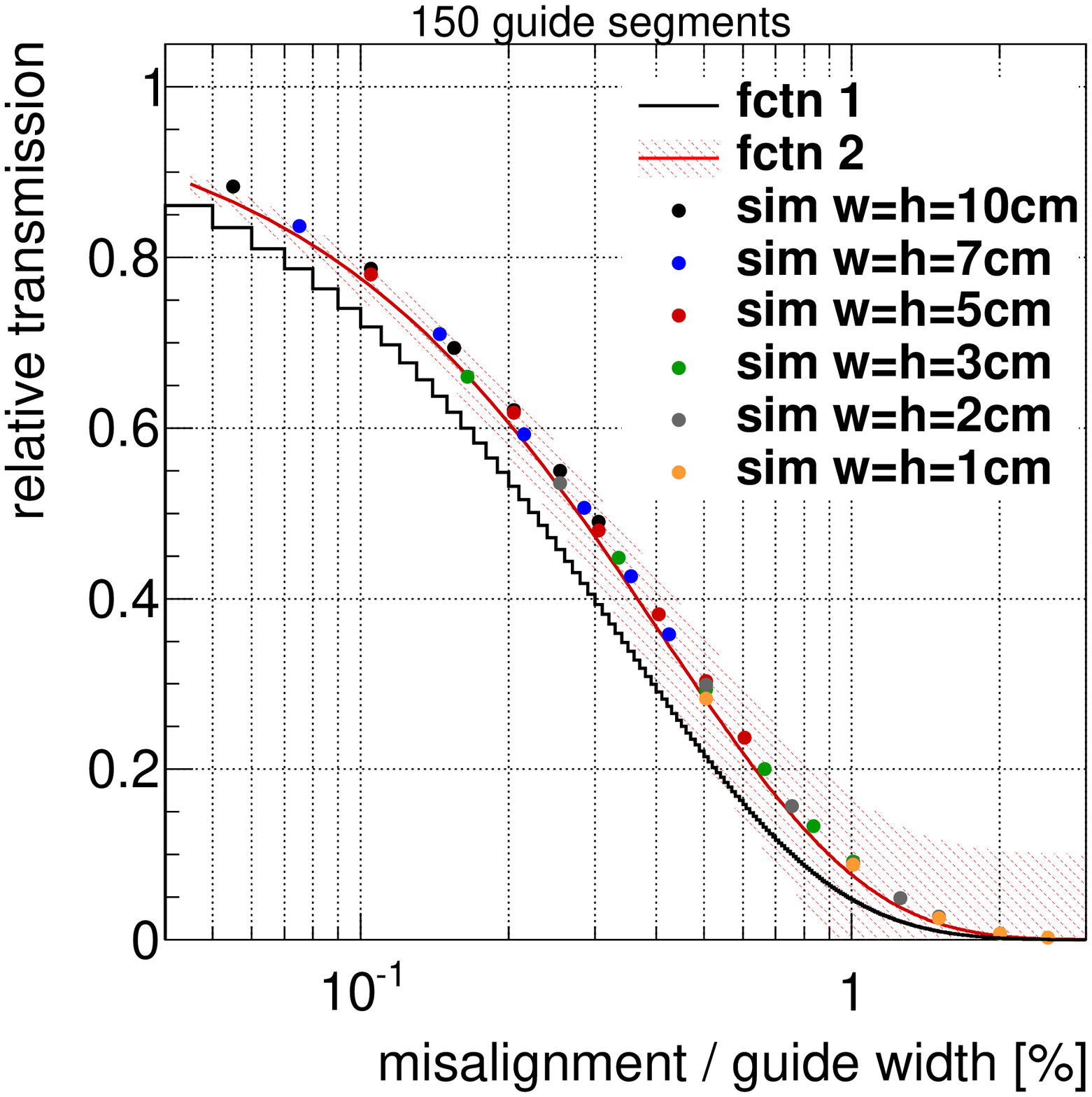}} 
\caption{Relative transmission of a (a) 20\,m and a (b) 150\,m guide
  with misalignment compared to the transmission calculated with the
  adjusted function~\ref{eq_fctn2}, showing that the improved model
  accurately accounts for the expected misalignments over long guides
  at the ESS.}
\label{f_Theory2}
\end{figure}

For long neutron guides made from many segments, the function
systematically overestimates beam losses due to neglect of the
gaussian nature of the spatial uncertainty caused by the misalignment:
68\% of the joints will be better aligned than the assumed value.

Equation~\ref{eq_fctn1} can be extended to reflect this effect:
\begin{equation}\label{eq_fctn2}
T=\left( 1-\frac{\delta w}{w}-\frac{\delta h}{h} \right)^{N} \cdot (1+\frac{N}{w}\frac{f}{100}\frac{\delta}{50\mu m})
\end{equation}
where $f=0.20 \pm 0.02$ is the percent difference between gaussian and
fixed misalignment per guide segment, per 50\,$\mu m$ misalignment and
per guide width (in cm), extracted from 20 simulations with different
random numbers for a 20\,m long 3$\times$3\,cm\sq\,guide and a
gaussian fit to the resulting distribution of $f$-values. This new
function is also shown in figure~\ref{f_Theory2}, with a 10\% error
band (assuming the uncertainty of $f$ to be the dominant one). It fits
the simulated data much better than function~\ref{eq_fctn1}.


\subsubsection{Transmission loss from misalignment in curved guides} \label{sec_Curved1}

The 20\,m long guide from the previous paragraph is now curved by 19
kinks at the joints of the 1\,m long guide segments, with curvature
radii between 1600\,m and 160\,m. No difference in beamloss is
observed, even for a large curvature with small $R$. Note, however,
that in a real guide, a stronger curvature would require shorter guide
segments, potentially increasing the beamlosses at the joins.

\begin{figure}[!hbt]
\subfigure[Ballistic guide. \label{f_CurvedBallisticB}]{\includegraphics[width=0.9\linewidth]{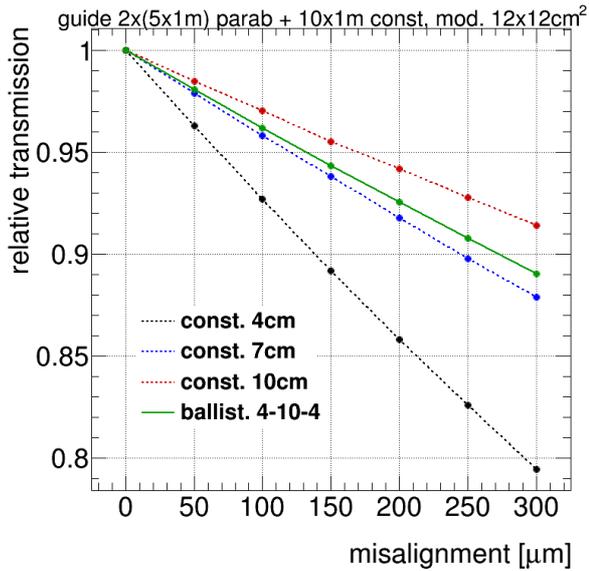}}
\subfigure[Multi-channel straight guide. \label{f_Bender}]{\includegraphics[width=0.9\linewidth]{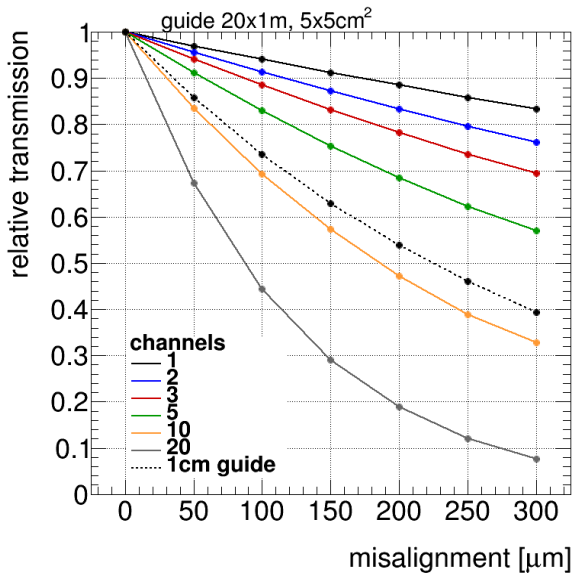}}
\caption{Relative transmission as a function of misalignment for a
  20\,m long guide consisting of 20 pieces. Random misalignment is modelled at
  each joint for 50\,$\mu$m, and scaled to larger
  misalignment. Solid lines are guides for the eye. (a) Ballistic guides of given widths, as stated in legend. (b) Multi-channel straight
  guides with different numbers of channels, as stated in legend, with a 1 cm square guide as a reference (dashed line).}
\label{f_CurvedBallistic}
\end{figure}

\subsubsection{Transmission loss from misalignment in ballistic guides}\label{sec_Ballistic1}

Figure~\ref{f_CurvedBallisticB} shows the guide loss with a ballistic
guide which contains two parabolic 5\,m long (de)focusing sections
with equal width and height between 4\,cm and 10\,cm, as well as a
central 10\,m long 10\,cm wide constant guide. The segment length is
again 1\,m (i.e. the parabolic shape is somewhat crude). For
comparison, the beamloss of constant guides with width and height
equal to the maximum, minimum and medium height are displayed
again. As one might expect, a ballistic guide shows losses slightly
exceeding those of a medium dimensioned guide with constant
cross section.

\subsubsection{Transmission loss from misalignment in multi-channel neutron guides}
In case of a multi-channel bender, the effect of misalignment on the
transmission can be expected to be comparable to the one of a guide
with a cross-section equal to one channel, plus an additional loss due
to the mis-match of the channel-separating blades.  For a fixed
substrate thickness of 0.5\,mm\footnote{motivated by consultation of a
  possible vendor}, the relative transmission for a
5$\times$5\,cm\sq\,guide with several channels is shown in
figure~\ref{f_Bender}.  The guide is not curved in this example (even
though a multi-channel neutron guide is usually used in combination
with curvature) but, since the primary loss mechanism is the mismatch
at the entrance planes, the losses as shown here should be
representative.

For comparison, a 1\,cm guide is shown again (dotted line): the
5-channel guide (green solid line) shows a slightly higher beamloss
from misalignment than could be expected purely from its channel
dimensions of 1$\times$5\,cm\sq, which would place it halfway between
the 1$\times$1\,cm\sq\,(black dotted) and the
5$\times$5\,cm\sq\,(black solid) lines. The additional beamloss due to
blade overlaps increases with the amount of misalignment.

\subsection{Angular misalignment}

In this calculation, we consider the impact that angular misalignment
has on the transmission, angular misalignment that arises from errors
in spatial positioning from the previous sections, {\it i.e.}
$\delta$, by $\delta\alpha=\frac{\delta}{L_{piece}}$. As in the
previous section, the total guide length is 20\,m and the guide
section length is 1\,m. Rotations about all three possible axes are
considered in the simulation of a 3$\times$3\,cm\sq\,curved guide with
radius 1600\,m or 160\,m. The observed beamloss is negligible compared
to the loss caused by a corresponding spatial misalignment ($<$2\% for
300\,$\mu$m misalignment).

\subsection{Overillumination}

Overillumination is applied by expanding the dimensions of a guide
upstream of a gap or location of strong expected misalignment.  This
removes any gaps in the phase space.  This was modelled by expanding
the guide section by $2\times$ the misalignment parameter.

The result of these geometrical changes is a guide entrance that is
now slightly larger than the exit.  This can create issues at the beam
extraction, one must take care with the acceptance of the guide
entrance increasing.  Clearly, the dimensions required for the neutron
source can increase, and insufficient phase space at the guide
entrance can offset any gains in transmission efficiency.  Moreover,
if overillumination is applied in the horizontal plane, it can affect
the radius of curvature needed to block line of sight to the source.

\begin{figure}[!hbt]
\subfigure[20\,m guide \label{f_Overillumination20}]{\includegraphics[width=0.9\linewidth]{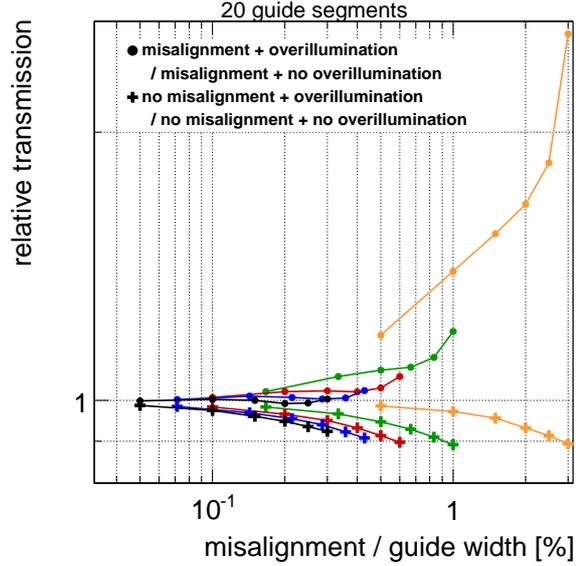}}
\subfigure[150\,m guide \label{f_Overillumination150}]{\includegraphics[width=0.9\linewidth]{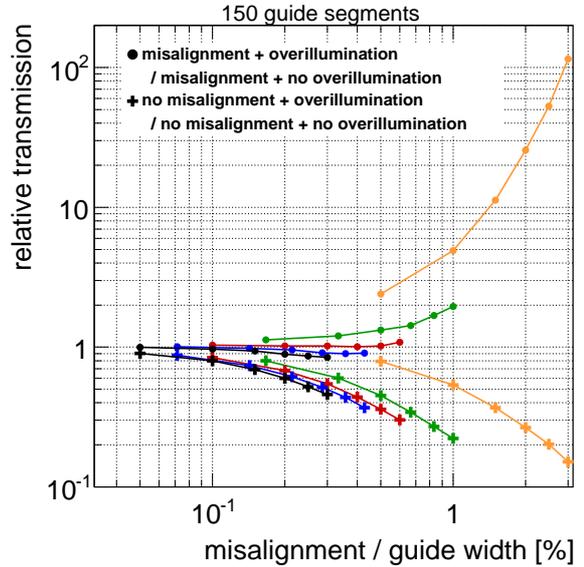}}
\caption{Gain in transmission provided by overillumination: relative transmission as a
  function of percental misalignment for a (a) 20\,m and a (b) 150\,m
  long guide consisting of 1\,m long segments. The source size is
  equal to the nominal guide width.
  The gain from overillumination
  when misalignment is present (round markers) compared to the case where misalignment is not present (crossed
  markers). Without misalignment, the given misalignment value merely
  determines the amount of misalignment anticipated in the design of the overillumination.
  The marker colour denotes the dimensions of the guide with the same scheme as for earlier figures (e.g. \ref{f_Theory1a})
  where black denotes: $w=h=10$ cm;
  blue: 7 cm;
  red: 5 cm;
  green: 3 cm;
  grey: 2 cm;
  orange: 1 cm.
  Note the double-logarithmic scale.}
\label{f_Overillumination}
\end{figure}

Omitting these adjustments here, and leaving the source size fixed to
be the same as the guide exit size (a fair evaluation bearing in mind
the ESS ``butterfly'' moderators are compact in the vertical
dimension) the effect of applying overillumination in order to prevent
a descreased transmission due to misalignment is investigated by
simply increasing the guide cross-section backwards for each section.
By so doing, we can examine if such a passive alignment strategy can
be applied across the whole system. Figure~\ref{f_Overillumination}
shows the gain that overillumination provides in the case where
misalignment is present (round markers) and where the guide has no
misalignment (crossed markers).

The beamloss can be largely reduced by overillumination for small
guide cross-sections (orange and green markers), while the mitigation
effect gets smaller with increasing guide width, such that for example
a 10\,cm wide guide (black markers) made of 20 pieces loses more
transmission by the overillumination approach if the guide is better
aligned than expected, than it would preserve in case of misalignment.
The losses here are caused by the aforementioned finite source size.

With an increasing number of guide segments, naturally both effects
increase: for large guide cross-sections, the loss caused by the
overillumination itself can become larger than the prevented
misalignment loss even when the guide segments are as poorly aligned
as expected. Applying the overillumination approach is hence not
constructive in such cases. Guides with small cross-section, on the
other hand, can gain even more from overillumination, so a detailed
risk benefit assessment for the expected misalignment should be
performed in each individual case.

For a fixed guide cross-section, overillumination becomes less
effective with more guide segments. In a long guide consisting of many
pieces, the overillumination approach should either be applied at a
view key positions only, or designed for a smaller misalignment than
expected.

In conclusion, overillumination should certainly not be used for large
numbers of elements if there is the possibility that the alignment
would be better than anticipated, otherwise it creates more losses
than it prevents.  One should probably only use this passive method in
one or two key locations where the movements are highly probable and
large, such as structural joins in the floor, and particularly for
guides with small cross sections.

\subsubsection{Effect of overillumination in curved guides}

In a curved guide, the effect of misalignment is the same as in a
straight guide (cf. section~\ref{sec_Curved1}). The loss caused by
overilluminating subsequent guide sections (without misalignment
really being present) is slightly larger than in a straight guide,
with the difference increasing with the expected misalignment. Hence
the overillumination method should only be applied in curved guides
with small cross-sections for which a significant loss from
misalignment is expected otherwise.